\documentclass[]{elsart}
\usepackage{epsfig}

\def\beq{\begin{equation}}
\def\eeq{\end{equation}}
\def\bea{\begin{eqnarray}}
\def\eea{\end{eqnarray}}

\newcommand{\twographs}[2]{%
\unitlength=1in
\begin{picture}(6,2.5)
\put(2.7,0){\epsfig{file=#2.eps, width=3.3in}}
\put(0,0){\epsfig{file=#1.eps, width=3.3in}}
\put(0,2.1){(a)}
\put(2.7,2.1){(b)}
\end{picture}}

\newcommand{\sixgraphs}[6]{%
\unitlength=1in
\begin{picture}(6,7.5)
\put(-0.6,5){\epsfig{file=#1.eps, width=3.7in}}
\put(0.33,5.47){\epsfig{file=#12.eps, width=2.509in}}
\put(2.3,5){\epsfig{file=#2.eps, width=3.7in}}
\put(3.23,5.47){\epsfig{file=#22.eps, width=2.509in}}
\put(-0.6,0){\epsfig{file=#5.eps, width=3.7in}}
\put(0.33,0.47){\epsfig{file=#52.eps, width=2.509in}}
\put(2.3,0){\epsfig{file=#6.eps, width=3.7in}}
\put(3.23,0.47){\epsfig{file=#62.eps, width=2.509in}}
\put(-0.6,2.5){\epsfig{file=#3.eps, width=3.7in}}
\put(0.33,2.97){\epsfig{file=#32.eps, width=2.509in}}
\put(2.3,2.5){\epsfig{file=#4.eps, width=3.7in}}
\put(3.23,2.97){\epsfig{file=#42.eps, width=2.509in}}
\put(0,4.6){(c)}
\put(0,7.1){(a)}
\put(2.9,7.1){(b)}
\put(0,2.1){(e)}
\put(2.9,2.1){(f)}
\put(2.9,4.6){(d)}
\end{picture}}

\newcommand{\modgraphs}[6]{%
\unitlength=1in
\begin{picture}(6,7.5)
\put(0,5){\epsfig{file=#1.eps, width=3.3in}}
\put(3,5){\epsfig{file=#2.eps, width=3.3in}}
\put(0,0){\epsfig{file=#5.eps, width=3.3in}}
\put(3,0){\epsfig{file=#6.eps, width=3.3in}}
\put(0,2.5){\epsfig{file=#3.eps, width=3.3in}}
\put(3,2.5){\epsfig{file=#4.eps, width=3.3in}}
\put(0,4.6){(c)}
\put(0,7.1){(a)}
\put(2.9,7.1){(b)}
\put(0,2.1){(e)}
\put(2.9,2.1){(f)}
\put(2.9,4.6){(d)}
\end{picture}}

\journal{Physics Letters}

\begin{document}
\hfill\parbox{8cm}{\raggedleft DAMTP-2006-5 \\ hep-ph/0601089 \\
}
\begin{frontmatter}
\title{Naturalness Priors and Fits to the Constrained Minimal Supersymmetric Standard Model}

\author[DAMTP]{B.C.  Allanach}

\address[DAMTP]{DAMTP, CMS, University of Cambridge, Cambridge, CB3 0FY, United Kingdom}

\begin{abstract}
We examine the effect of a prior that favours low values of fine-tuning on Bayesian multi-dimensional fits of the constrained minimal supersymmetric standard model (CMSSM or mSUGRA) to current data. The dark matter relic density, the anomalous magnetic moment of the muon and the branching ratio of $b \rightarrow s \gamma$ are all used to constrain the model via a Markov Chain Monte Carlo sampler.  
As a result of the naturalness prior, posterior probability distributions skew
towards lighter higgs and sparticle masses, the effect being most pronounced
in the gaugino sector. 
Interestingly, slepton masses are an exception and skew towards heavier
masses.  
The lightest CP-even Higgs $h^0$-pole annihilation mechanism becomes allowed at 
the 2$\sigma$ level for the latest combination of measurements of $m_t=172.7
\pm 2.9$ GeV, provided we allow for a theoretical error in the prediction of
its mass $m_{h^0}$.  $m_{h^0}$ is constrained to be less than 120 GeV at
the
95$\%$ C.L. Probing the branching ratio of $B_s \rightarrow \mu^+ \mu^-$ to the
level of 2$\times 10^{-8}$, as might be achieved by the Tevatron experiments,
would cover 32$\%$ of the probability density, irrespective of which of the two priors is used.
\end{abstract}
\end{frontmatter}

\section{Introduction}
Although weak-scale supersymmetry (SUSY)~\cite{haber} arguably offers the best candidate for physics beyond the Standard Model,
it potentially suffers from a fine-tuning problem~\cite{ftrefs}. Radiative electroweak symmetry breaking conditions imply that the mass of the $Z^0$-boson $M_Z$ is related to the superpotential $\mu$-term, the two soft supersymmetry breaking Higgs mass parameters $m_{H_{1,2}}$ and the ratio of the Higgs vacuum expectation values $v_2/v_1=\tan \beta$ by
\begin{equation}
\frac{M_Z^2}{2} = \frac{m_{H_1}^2 - m_{H_2}^2 \tan^2 \beta}{\tan^2 \beta - 1} -\mu^2   \label{rewsb}
\end{equation}
at tree level~\cite{haber}. In the following numerical calculations, the full one-loop MSSM corrections are added to this relation. Experiments at LEP2 and the Tevatron have placed increasingly severe 
bounds upon sparticle and Higgs masses~\cite{ftrefs,PDBook} 
and in constrained models they
push up the minimum value of $\mu$ to several hundred GeV. In 
order to reproduce the observed value of $M_Z=91$ GeV, there then must be a large
degree of cancellation between the {\em a priori} unrelated terms on the right
hand side of Eq.~\ref{rewsb}. This is considered to be unnatural. In order to
quantify the amount of fine-tuning, one can construct~\cite{ftrefs} a fine-tuning measure $c_i$ for a parameter of the model $p_i$ by noting that $M_Z$ is unstable to small changes of $p_i$ as a result of the fine-tuning: 
\begin{equation}
c_i \equiv \left|\frac{\partial \ln M_Z}{\partial \ln p_i}\right|, \qquad
c\equiv \mbox{max} \{c_i \}. \label{fineT}
\end{equation}
$c_i$
then measures the fractional change in $M_Z$ (the partial derivative is calculated while {\em not} using the constraint in Eq.~\ref{rewsb}) produced by some small fractional  change in the parameter $p_i$. 
We shall here use the largest $c_i$ to quantify the amount of fine-tuning $c$ for a given point in parameter space. 

It is our purpose in this paper to study the effect of a prior probability distribution that favours a low value of $c$ in combined fits to the CMSSM\@. We wish to extract any collider implications, as well as implications for the dark matter annihilation mechanism. 
The MSSM contains several good candidates for cold dark matter, the most studied being the lightest neutralino $\chi_1^0$. 
The assumption that thermally produced  neutralinos make up the dominant component of dark matter leads to strict constraints upon the parameter space of the MSSM\@. We shall here take this assumption, as well as the assumption of R-parity (which provides stability for the lightest neutralinos). In order to keep the dimensionality of the parameter space down, we shall further specialise to the CMSSM. The CMSSM unifies all SUSY breaking scalar mass parameters to $m_0$, the Standard Model gaugino mass parameters to $M_{1/2}$ and the SUSY breaking trilinear scalar couplings to $A_0$ at a high energy scale, commonly taken to be the scale of electroweak gauge unification $M_{GUT}$.  Our choice of parameters from which to calculate the fine-tuning includes the SUSY breaking Higgs bilinear $B$: 
\begin{equation}
p_i \in \{ m_0, A_0, M_{1/2}, B, \mu \}.
\end{equation}
$B$ is used in preference to $\tan \beta$ since it directly appears in the Lagrangian and is therefore considered to be more fundamental. 
Our definition of $c$ does not include the sensitivity to the GUT-scale top Yukawa coupling $h_t$. We can argue about whether to include it or not in the definition since its value is determined by aspects of flavour physics in the model, an orthogonal aspect to the SUSY breaking aspects studied in the present paper. Were we to include it in the definition of $c$, a strong suppression of the high $m_0$ regime would be the likely result of the naturalness prior, since $c_{h_t}$ is very high there~\cite{Feng:1999mn}. 

In ref.~\cite{Giusti:1998gz}, exclusion limits from LEP1 and LEP2 were placed
upon the CMSSM and a probability of $1/c$ was applied to the remaining
parameter space. This resulted in probability distributions for sparticle
masses and various observables. This work is similar in spirit to the
ingredient of the naturalness prior that we will follow.
A more common approach in the literature involves applying separate 95$\%$ C.L. constraints upon a two-dimensional sub-space of the model, see recent examples in ref.~\cite{classic}. The advantage of this simple approach is that it is easy to discern which constraints act on different parts of parameter space. On the other hand, it was argued in Ref.~\cite{myold} that a more complete approach would perform a combined likelihood fit in the full dimensionality of the parameter space, including variations of important Standard Model input parameters. 
In fact, the feasibility of the approach had already been demonstrated in a four-dimensional subspace~\cite{gondolo}.
This was achieved by utilising the Metropolis algorithm in a Markov Chain Monte Carlo (MCMC) algorithm~\cite{mackay}. The MCMC approach to parameter sampling has a calculation time that scales linearly with the number of dimensions, rendering it very useful for investigating parameter spaces with dimensionality greater than 3. We point the interested reader to refs.~\cite{myold,mackay} for details on the algorithm, but the important point for the present paper is that the algorithm provides a list (or ``chain'') of points in parameter space. The density of points that it produces is proportional to some specified function of the parameters. In reference~\cite{myold}, the function chosen was the likelihood, the probability distribution function (pdf) of reproducing data $d$ given a point in CMSSM parameter space $m$: ${\mathcal L}=p(d|m)$. One ideally wishes to determine the posterior pdf that the model point $m$ is correct given the data $d$: $P(m)=p(m|d)$. The relative posterior between
two different points $m_{1,2}$ in parameter space can be determined provided one assumes a prior pdf for the model parameters $p(m)$:
\begin{equation}
\frac{P(m_1)}{P(m_2)} = \frac{p(d|m_1)}{p(d|m_2)} . \frac{p(m_1)}{p(m_2)}.
\end{equation}
Thus, the likelihood distribution is proportional to the posterior distribution for the case of flat prior distributions, i.e. when $p(m_1)/p(m_2)$ is a constant across the relevant parameter space. For a brief discussion on a Bayesian interpretation of the prior probability distributions, see the Appendix.

Here, we will go beyond the MCMC analysis of Ref.\cite{myold} by using a
plausible {\em naturalness}-favouring prior proportional to $1/c$. This will
then disfavour regions of parameter space with high fine-tuning, arguably a
commendable effect. 
We will then compare and contrast the resulting probability distributions of
observables, parameters and sparticle masses with the flat-prior case. 

Having obtained a chain of points whose density is proportional to their
posterior assuming some prior, we could in principle just re-weight the
probability of each point
by some new prior and re-plot all of the results. In the present case, we
would for example use the flat-prior sample, but reweight the probability of
each point by $1/c$, 
then re-bin any results. The only disadvantage to this approach is that if
previously unlikely regions of
parameter space now have high probabilities, they will not have been sampled
very often in the initial Markov chain run. Thus the statistical fluctuations
in this region will be large. If the prior doesn't make a lot of difference to
the posterior, this degradation of the statistics may not matter much. Here
though, we will find a significant difference in the results due to different
priors, implying that a simple re-weighting procedure would suffer from poor
statistics. We have actually re-run the MCMC with the new prior, which
circumvents any potential statistics problem.

\section{Numerical Methodology}
The numerical analysis follows Ref.~\cite{myold} identically except for the inclusion of a naturalness prior and updates in the codes that calculate observables. We briefly include the main features of the analysis here, but we refer the interested reader to Ref.~\cite{myold} for more detail. 

The range of CMSSM parameters considered is 
sign$(\mu)=+1$, 
$A_0=$-2 TeV to 2 TeV,
$m_0=$ 60 GeV to 2 TeV,
$M_{1/2}=$60 GeV to 2 TeV and
$\tan \beta=$ 2 to 60.
 Although this range was initially considered on perturbativity and naturalness grounds, there is a further pragmatic region for not extending it, namely that an enlarged space will result in less efficiency in the MCMC algorithm. The MCMC algorithm already uses 4 weeks of CPU time on the {\tt lxplus} CERN cluster, rendering an enlarged run inconvenient. Note that recently, imaginary $m_0$ has received some attention in the literature~\cite{Feng:2005ba}. This region predicts a non-neutralino dark matter candidate and is not covered by the present analysis.

We use the following measurements in order to constrain the CMSSM: 
the running bottom quark mass in the minimal subtraction scheme $m_b (m_b)^{\overline{MS}} = 4.2 \pm 0.2 \mbox{~GeV}$~\cite{PDBook},  the top pole mass $m_t=172.7 \pm 2.9$ GeV~\cite{Group:2005cc}, 
the strong coupling constant in the modified minimal subtraction
scheme at $M_Z$
$\alpha_s (M_Z)^{\overline{MS}} = 0.1187 \pm 0.002$~\cite{PDBook}, the new physics contribution to the anomalous magnetic moment of the muon 
$\delta (g-2)_\mu/2 = 19.0 \pm 8.4 \times 10^{-10}$~\cite{Bennett:2004pv,Allanach:2005yq}, the branching ratio $BR(b \rightarrow s \gamma) = 3.52 \pm 0.42 \times 10^{-4}$~\cite{Gambino:2004mv} and the relic density of thermal dark matter $\Omega_{DM} h^2 = 0.1126^{+0.0081}_{-0.0091}$~\cite{Spergel:2003cb}.

For each parameter point, {\tt SOFTSUSY2.0.4}~\cite{Allanach:2001kg} was used to calculate the sparticle spectrum. The spectrum was linked via the SUSY Les Houches Accord~\cite{Skands:2003cj} to {\tt micrOMEGAs1.3.6}~\cite{Belanger:2001fz}, which computed predictions for $\Omega_{DM} h^2$, $BR(b \rightarrow s \gamma)$ and 
$\delta (g-2)_\mu/2$. Empirically derived cuts were then made on sparticle and Higgs masses~\cite{myold} from search data and the requirement of a neutralino lightest supersymmetric particle. 
A $\chi^2$ value $\chi_i^2$ was found for each of the six observables mentioned above, and the (un-normalised) posterior probability density is calculated as
\begin{equation}
P = \frac{1}{c}\prod_{i=1}^6 \mbox{exp}(-\chi^2_i/2),
\end{equation}
where $c$ is defined in Eq.~\ref{fineT}. The MCMC algorithm was then employed to provide a sampling of points in the 7 dimensional parameter space $m_0$, $M_{1/2}$, $A_0$, $\tan \beta$, $m_b(m_b)^{\overline MS}, m_t, \alpha_s(M_Z)$. The density of the resulting parameter space points is proportional to $P$ if the chain has converged.

In order to test the robustness of the numerical results, a convergence test was applied. As described in Ref.~\cite{myold}, 9 independent 
Markov chains of 10$^6$ potential points each were run, all with different
random number seeds and at different random starting points. An estimation of
the convergence of the individual chains can be determined by  comparing how
similar the distributions of interesting quantities are between the different
chains. The technique (invented in Ref.~\cite{stat}) as implemented in
Ref.~\cite{myold}, provides an estimated upper bound on the potential root
variance reduction of each scalar quantity if the chain were run for an
infinite number of steps. After 10$^6$ steps, the upper bound is less than the
required 5$\%$ for the naturalness prior sample and for the flat prior sample,
for every scalar distribution that we are interested in. The chains were
therefore deemed to have converged. For all of the numerical results presented
below, the full 9$\times$10$^6$ potential-point sample was used.  

\section{Posterior Probability Distributions}

\begin{figure}
\begin{center}
\sixgraphs{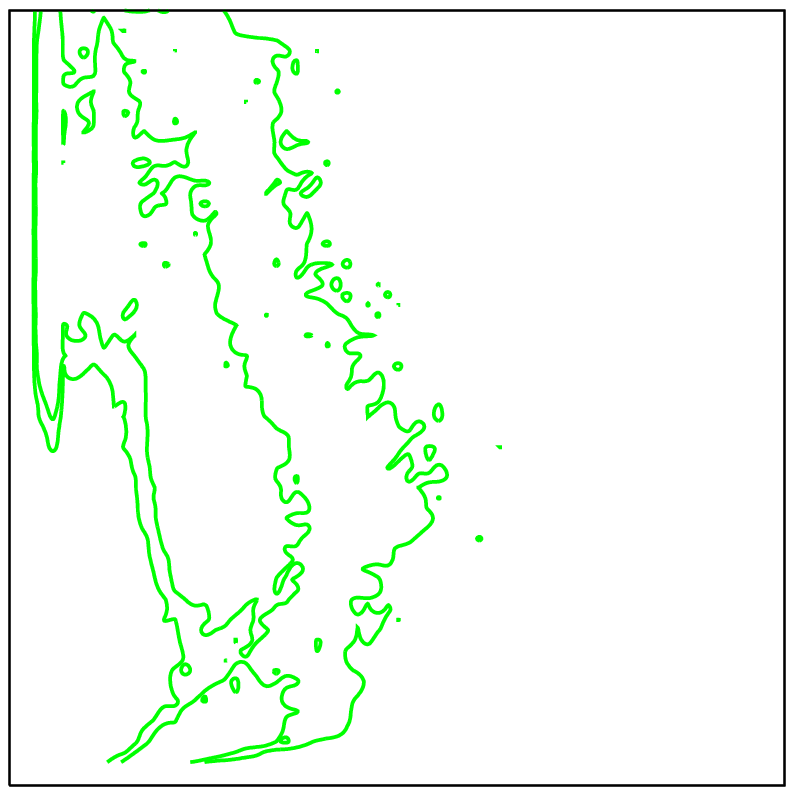}{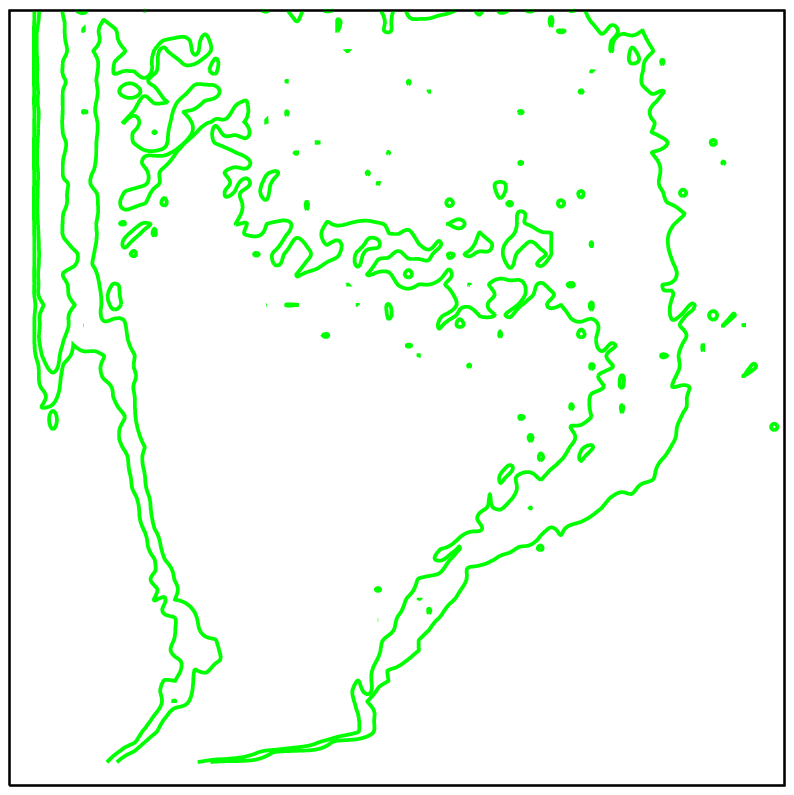}{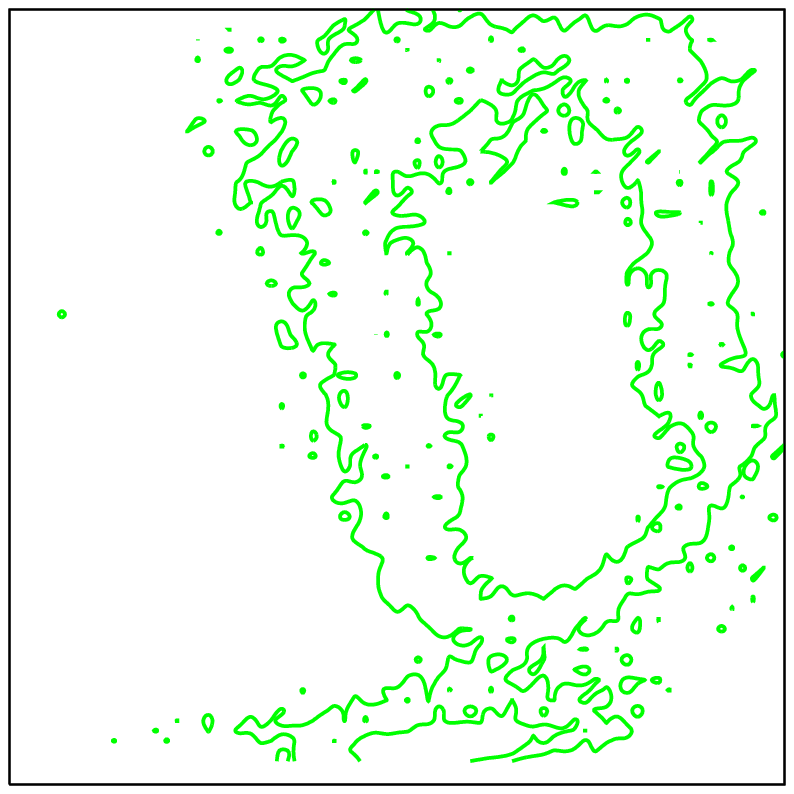}{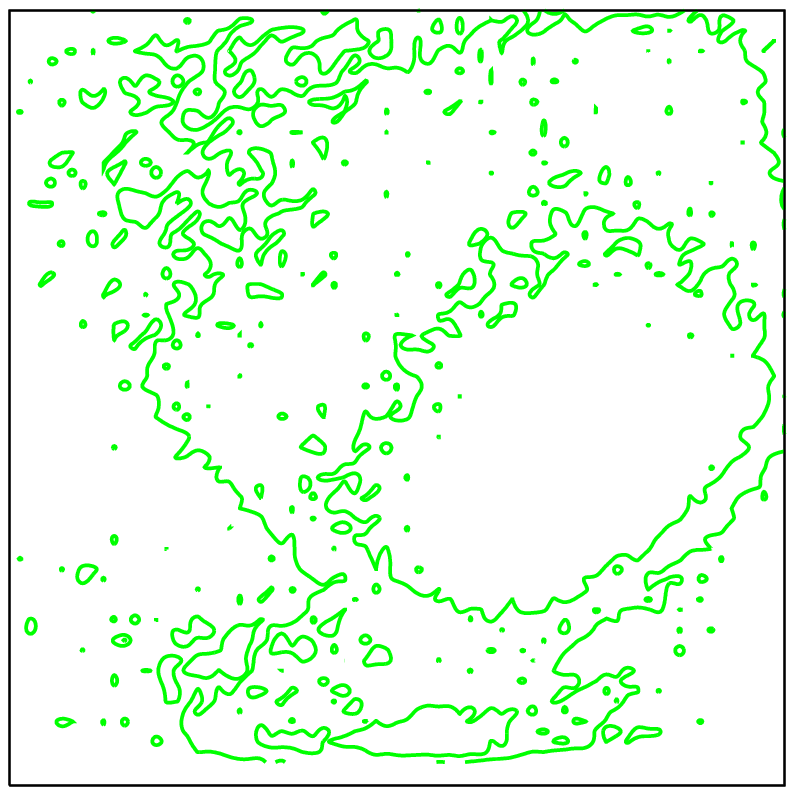}{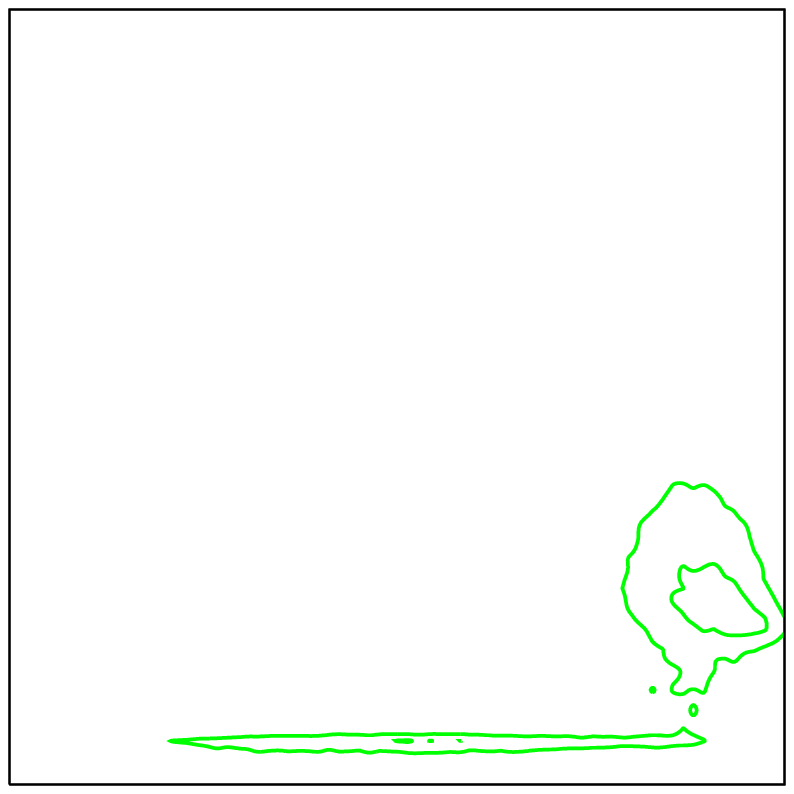}{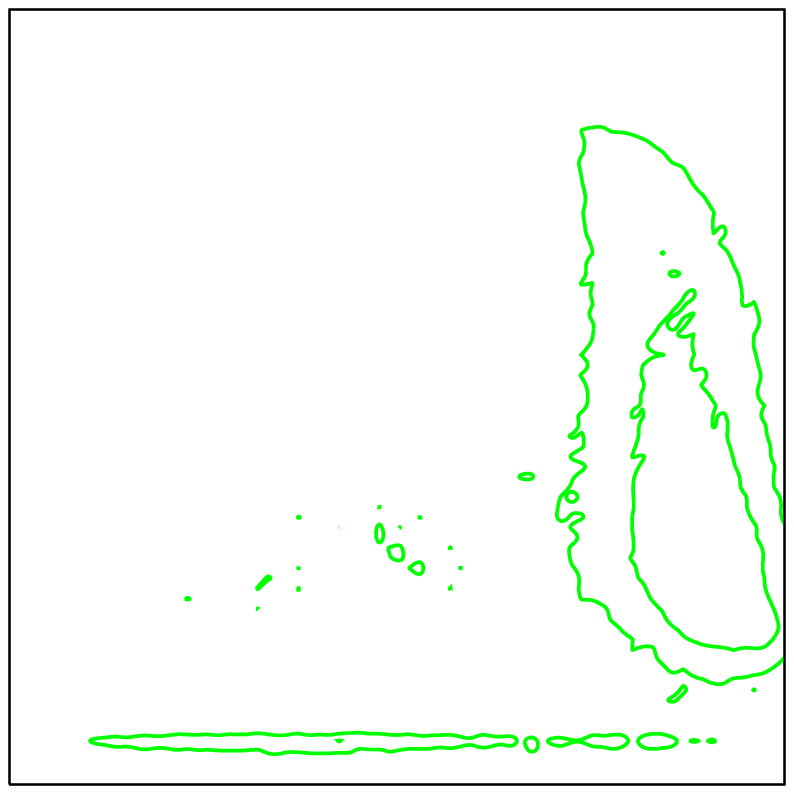}
\caption{Comparing constraints on the CMSSM parameter space assuming a
  naturalness prior in the left-hand plots and a flat prior
in the right-hand plots. Unseen CMSSM parameters have been marginalised over.
The posterior probability $P$ normalised to the maximum probability bin of
  each plot is determined by reference to the bar on the right hand side. The
  contours display the 67$\%$ and 95$\%$ C.L. contours respectively.} 
 \label{fig:planes}
 \end{center}
 \end{figure}

In Fig.~\ref{fig:planes}, we show the posterior distribution across various planes in parameter space. Unlike ref.~\cite{myold}, we have not shown all possible planes in the mSUGRA parameter space, but have instead concentrated on the planes from which the effects of the naturalness prior may be deduced. In every plane, the posterior probability distribution is shown, {\em marginalised} over the other five mSUGRA/Standard Model parameters. Marginalisation simply means that the unseen dimensions are integrated over. 
Individual MCMC points were placed in 75 by 75 bins in the 2 dimensional
space, and for each plot the probabilities are normalised by the probability
in the highest bin. Thus, one can consider the marginalisation process to be
an averaging of the posterior in the unseen dimensions. This can be viewed
simply as a way of visualising constraints in the 7 dimensional parameter
space, or alternatively as a probability distribution in the relevant plane,
using flat priors in the unseen dimensions. 

On all plots on the left hand side of Fig.~\ref{fig:planes}, the naturalness
prior has been used in order to calculate the posterior probability
distribution. For those on the right hand side, a flat prior has been used for
the purposes of comparison. The results with a flat prior have already
appeared in ref.~\cite{myold} for a previous version of {\tt SOFTSUSY}, and
are identical by eye to those shown here. 
As often pointed out in the literature, the CMSSM predicts too much dark matter relic density compared to observations unless a specific efficient annihilation mechanism acts.

In Fig.~\ref{fig:planes}b, the light Higgs pole region~\cite{Drees:1992am,Djouadi:2005dz} is the bright vertical sliver at the extreme left hand side of the plot (i.e. low $M_{1/2}$ and high $m_0$). It corresponds to efficient annihilation through an $s$-channel $h^0$. 
The stau co-annihilation region~\cite{Griest:1990kh}, where ${\tilde \tau}_1 \chi^0_1 \rightarrow \gamma \tau$, corresponds roughly to the bright region at lowest $m_0$, where the lightest stau ${\tilde \tau}_1$  and the $\chi_1^0$ are quasi mass degenerate.
The rest of the probability bulk contains the $A^0-$pole region~\cite{Drees:1992am} where $\chi_1^0$ pairs are close in mass to an $s-$channel $A^0$ resonance.
Some of the probability density corresponds to the simultaneous action of several of these channels, sometimes including the so-called focus point regime~\cite{Feng:1999mn} where the lightest neutralino contains a significant higgsino component and so $\chi_1^0$ pairs efficiently annihilate via gauge couplings to pairs of gauge bosons. Slepton co-annihilation also contributes in some regions~\cite{us}.

Comparing Figs.~\ref{fig:planes}a,b, we see that the $A^0$-pole region and stau co-annihilation regions have become less probable as a result of the naturalness prior. Also, we infer that higher values of $M_{1/2}$ have a penalty originating from the prior distribution. 
Comparing Figs.~\ref{fig:planes}c,d, we see the migration of the probability distribution away from low $m_0$ values toward higher values as a result of the naturalness prior. The local maximum in Fig.~\ref{fig:planes}c at $m_0 \sim 0.7$ TeV corresponds to the $A^0-$pole annihilation region, which also appears in Fig.~\ref{fig:planes}d. Another local maximum at $m_0 \sim 1.2$ TeV corresponds mainly to the increased $h^0-$pole region. The third maximum at the lowest values of $m_0$ decreases in probability as a result of the fine-tuning prior. 
A comparison of Figs.~\ref{fig:planes}e,f demonstrates again the migration towards lower $M_{1/2}$ as a result of the naturalness prior. It also becomes clear that the $h^0$-pole region (contained in the horizontal stripe at the bottom of both figures) increases in relative importance by using the naturalness prior. High $\tan \beta$ is somewhat preferred in both cases.

Although we have discussed the different regions corresponding to various annihilation mechanisms, the true situation is slightly less clearly defined. In order   to illustrate this, we use strict definitions of the different annihilation regions. We define the $h^0$ or $A^0-$pole regions by $|2 m_{\chi_1^0} / m_{h^0/A^0} - 1|<0.1$ and the stau co-annihilation region is defined by $|m_{\tilde \tau_1} / m_{\chi_1^0}-1|<0.1$. Under these definitions, the relative probabilities of the annihilation mechanisms are shown in Table~\ref{tab:mechs} for each different prior.
\begin{table}
\begin{center}
\begin{tabular}{|c|cc|}\hline
mechanism & flat prior & natural prior \\ \hline
$h^0-$pole & 0.025 & 0.07 \\
$A^0-$pole & 0.41 & 0.14 \\
${\tilde \tau}-$co-annihilation & 0.26 & 0.18\\
rest & 0.31 & 0.61\\ \hline
\end{tabular}
\caption{Relative posterior probabilities for the different co-annihilation regions, as defined in the text. The mechanism denoted ``rest'' indicates the probability of points not fitting into one of the clearly defined regions.\label{tab:mechs}} 
\end{center}
\end{table}
We see from the table that the naturalness prior does indeed increase the $h^0-$pole probability at the expense of the $A^0$-pole and ${\tilde \tau}$ co-annihilation regions, but the dominant effect is that most of the probability density lies outside the clearly defined regions in the case of the naturalness prior. 
The probability of the $h^0-$pole region is dependent upon theoretical errors assumed in the {\tt SOFTSUSY2.0.4} prediction for $m_{h^0}$. The LEP2 limits
have been softened by 3 GeV in order to take a 3 GeV uncertainty~\cite{Degrassi:2002fi} in the prediction into account in the manner described in Ref.~\cite{myold}. If we instead assume that the {\tt SOFTSUSY2.0.4} prediction of $m_{h^0}$ has no theoretical error, the $h^0$-pole region obtains probabilities of $2\%$ and 5$\%$ for the flat and naturalness prior cases respectively.
Investigation of the points in the ``rest'' region shows several competing
annihilation mechanisms at work at one point: i.e. different combinations of
bulk/focus-point/$A^0/h^0$-pole annihilation mechanisms. For example, the
best-fit point in the ``rest'' region is $m_0=1419$ GeV, $M_{1/2}=180$ GeV,
$A_0=550$ GeV, $\tan \beta=51.8$, $m_b(m_b)=4.20$ GeV, $m_t=172.4$ GeV,
$\alpha_s(M_Z)=0.1188$. This point results in a fine-tuning parameter of 17.7, 
$BR(b \rightarrow s \gamma)=3.15\times 10^{-4}$, $\delta (g-2)_\mu/2=10.9
\times 10^{-10}$, $\Omega_{DM} h^2=0.1096$, 
indicating a good fit to the data\footnote{A good fit is not surprising since
  we have one more parameter than observables used to form the likelihood.}
 at moderate values of the fine-tuning. 
Investigation of the annihilation mechanism at this point shows that 
annihilation into light quark-anti-quark pairs provides 8$\%$ of the
annihilation, $b \bar b$ provides 80$\%$ and $\tau \bar \tau$ 10$\%$. This is
consistent with non-negligible contributions from both the $h^0$ pole and the
$Z^0$ pole. The neutralino has a non-negligible higgsino component, so the
point can also be described as being in the focus-point regime.

\begin{figure}
\begin{center}
\unitlength=1in
\begin{picture}(4,3)
\put(0,0){\epsfig{file=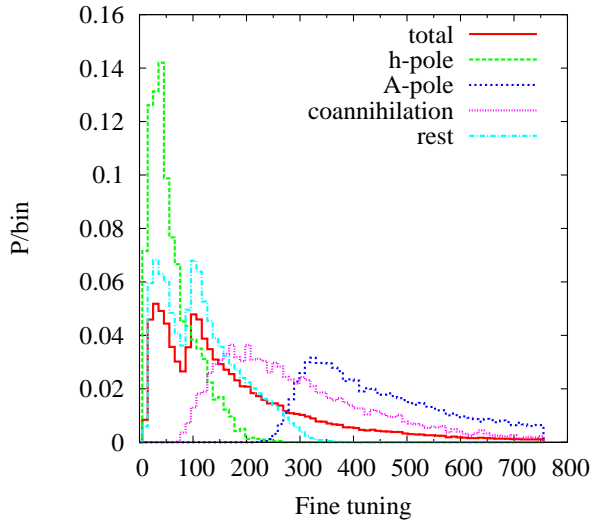, width=4in}}
\end{picture}
\caption{Fine-tuning probability distribution $c$ of various annihilation
  mechanisms. Each individual histogram has been normalised to an integrated
  probability of 1.} 
 \label{fig:ft}
 \end{center}
 \end{figure}
Fig.~\ref{fig:ft} shows probability distributions for the different annihilation mechanisms. We have fixed the normalisation of each histogram such that its integrated probability is 1. The $1/c$ tendency of each histogram due to the naturalness prior is evident at large $c$. The $h^0-$pole is clearly favoured by the naturalness prior, whereas stau co-annihilation and $A^0-$poles show higher values of the fine-tuning and thus become relatively disfavoured, corroborating our interpretation of Fig.~\ref{fig:planes}. A similar investigation of the different components of $BR(b \rightarrow s \gamma)$ and $(g-2)_\mu$ shows much less difference between the distributions of the different annihilation regions. 
Further investigation of the MCMC samples finds that
13$\%$ of the $A^0$ pole region is contaminated with points that also satisfy
the $\tilde \tau$-co-annihilation constraint.

\begin{figure}
\begin{center}
\twographs{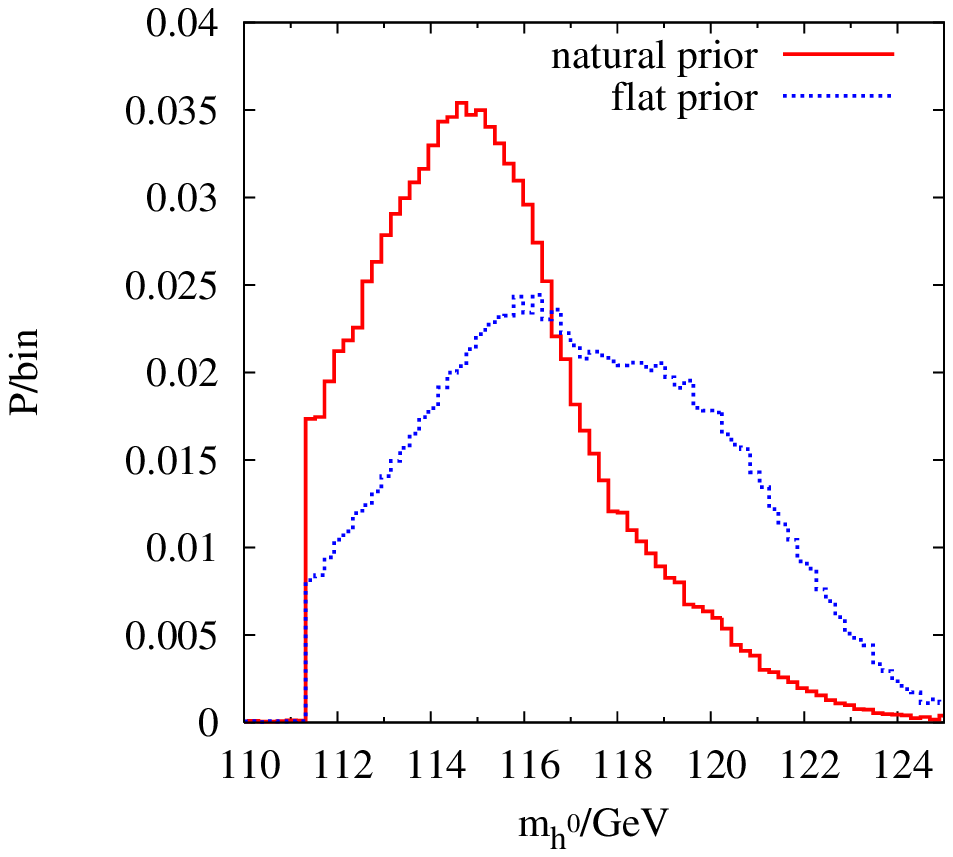}{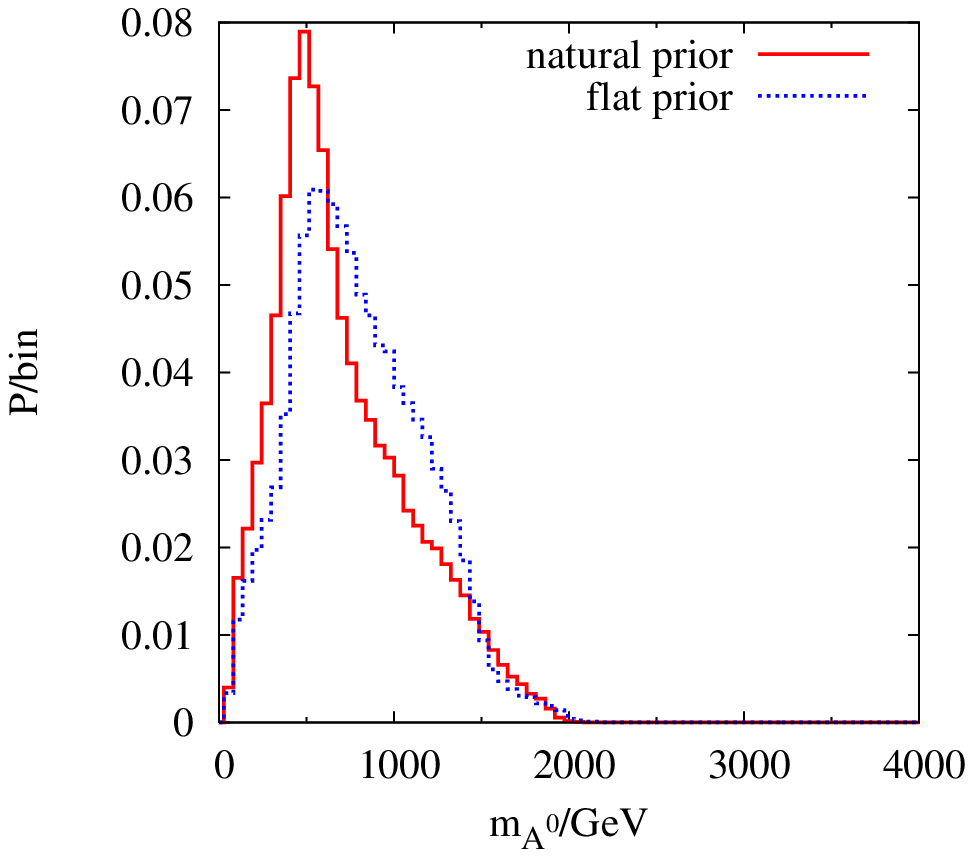}
\caption{Comparison of mass probability distributions for higgs bosons with (``natural prior'') and without (``flat prior'') the naturalness prior.
Each histogram has been normalised to an integrated probability of 1. 
The plots show distributions for, respectively, (a) the lightest CP-even higgs mass and (b) the CP-odd higgs mass.} 
 \label{fig:hMasses}
 \end{center}
 \end{figure}
Probability distributions for $m_{h^0}$ and $m_{A^0}$ are shown in Fig.~\ref{fig:hMasses}. The naturalness prior skews the $h^0$ distribution toward lighter masses. The $A^0$ distribution shape also skews to lower values, except for a small enhanced tail at the heaviest masses. Upper 95$\%$ C.L. limits on $h^0$ and $A^0$ masses are displayed in Table~\ref{tab:conf}.

\begin{table}
\begin{center}
\begin{tabular}{|c|cc|} \hline
particle & flat prior & natural prior  \\ \hline
$h^0$ &0.123 & 0.120\\
$A^0$ & 1.45& 1.50\\
$\chi_1^0$ &0.65 & 0.45\\
$\chi_1^\pm$ &1.20 & 0.85\\
${\tilde g}$ &3.25 & 2.30\\
${\tilde e}_R$ & 1.90& 1.90\\
${\tilde q}_L$ &3.20 & 2.45\\
${\tilde t}_1$ &2.45 & 1.80\\
\hline
\end{tabular}
\end{center}
\caption{Upper 95$\%$ C.L. limits on various MSSM particle masses from the fits. Particle masses are shown in units of TeV and, except for the lightest CP-even higgs $h^0$, have been rounded to the nearest 50 GeV.\label{tab:conf}}
\end{table}

\begin{figure}
\begin{center}
\modgraphs{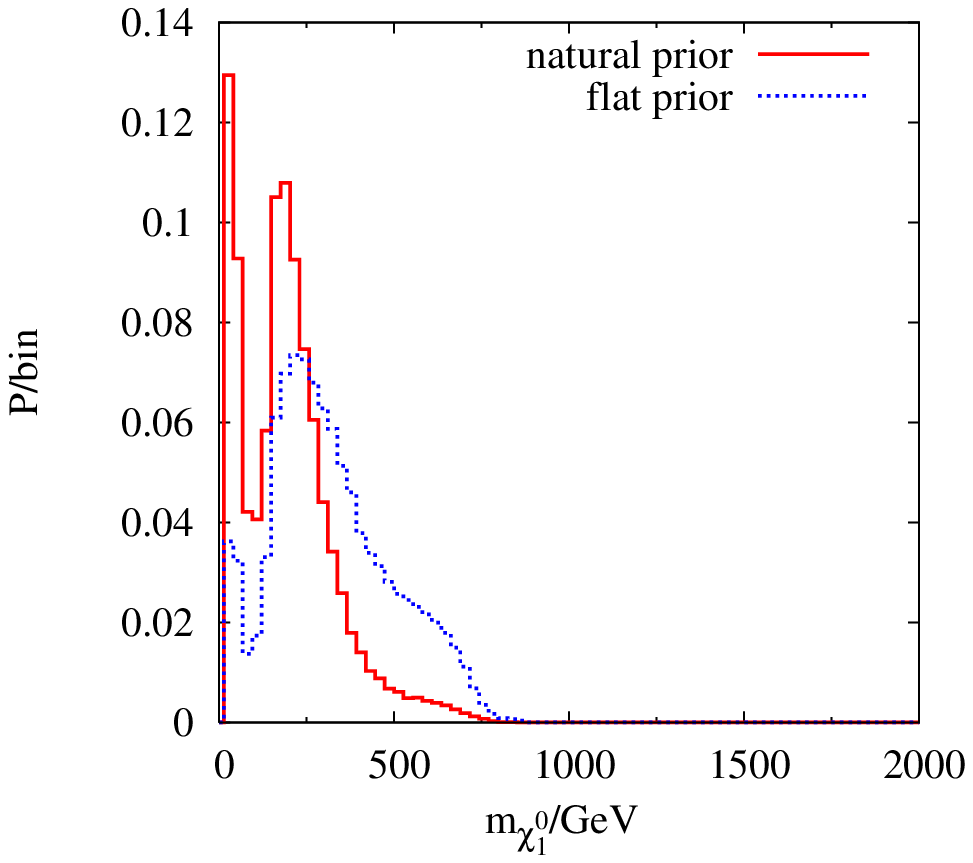}{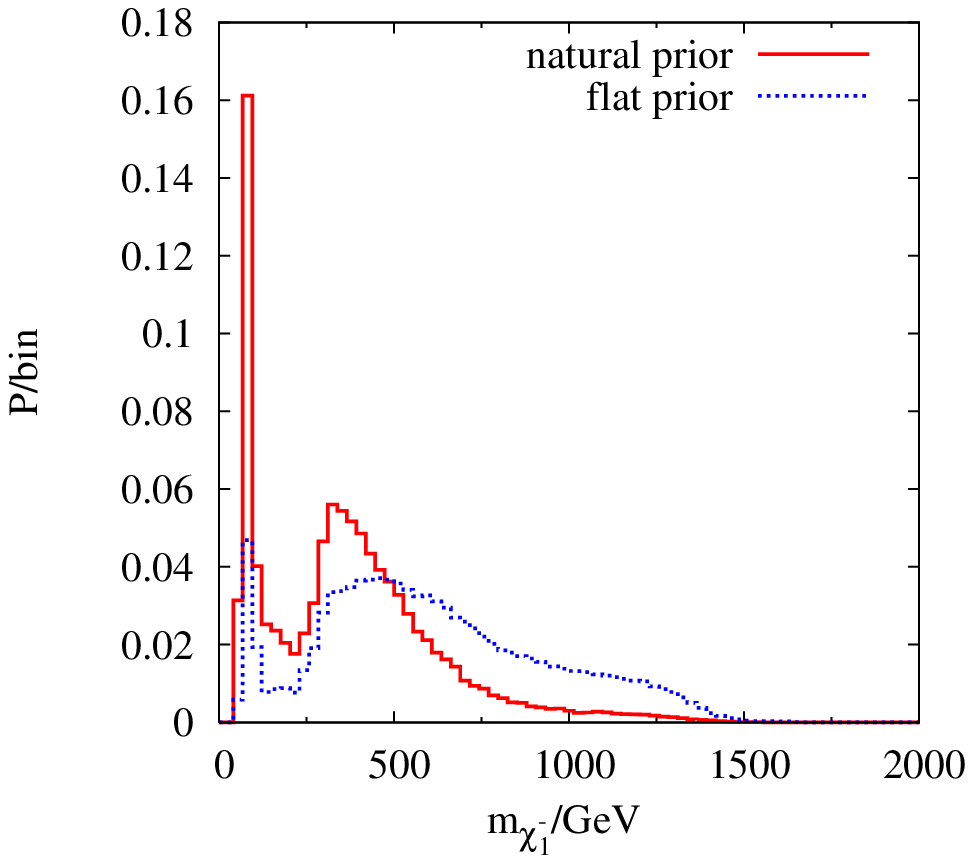}{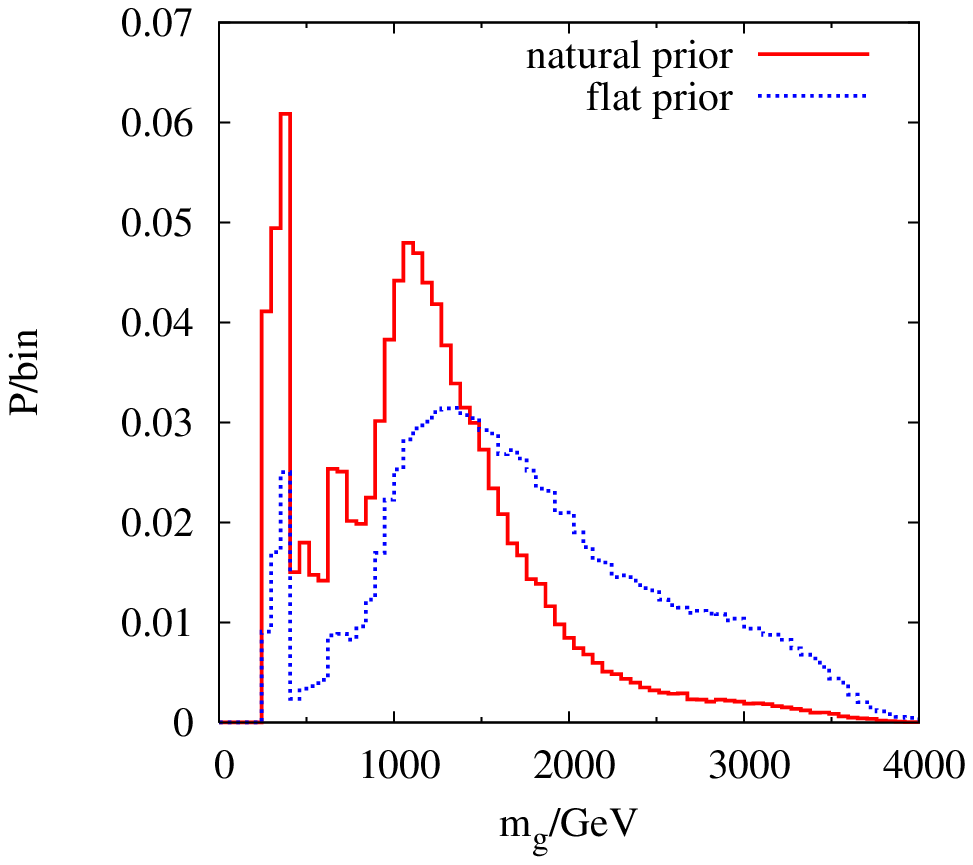}{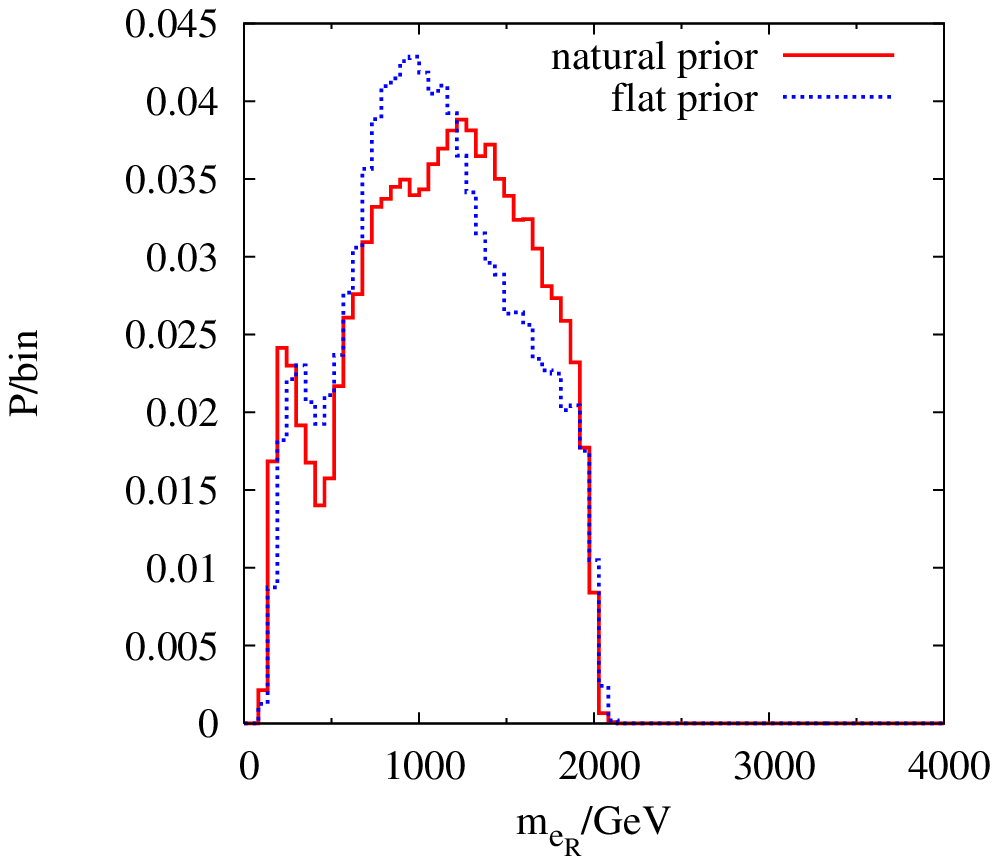}{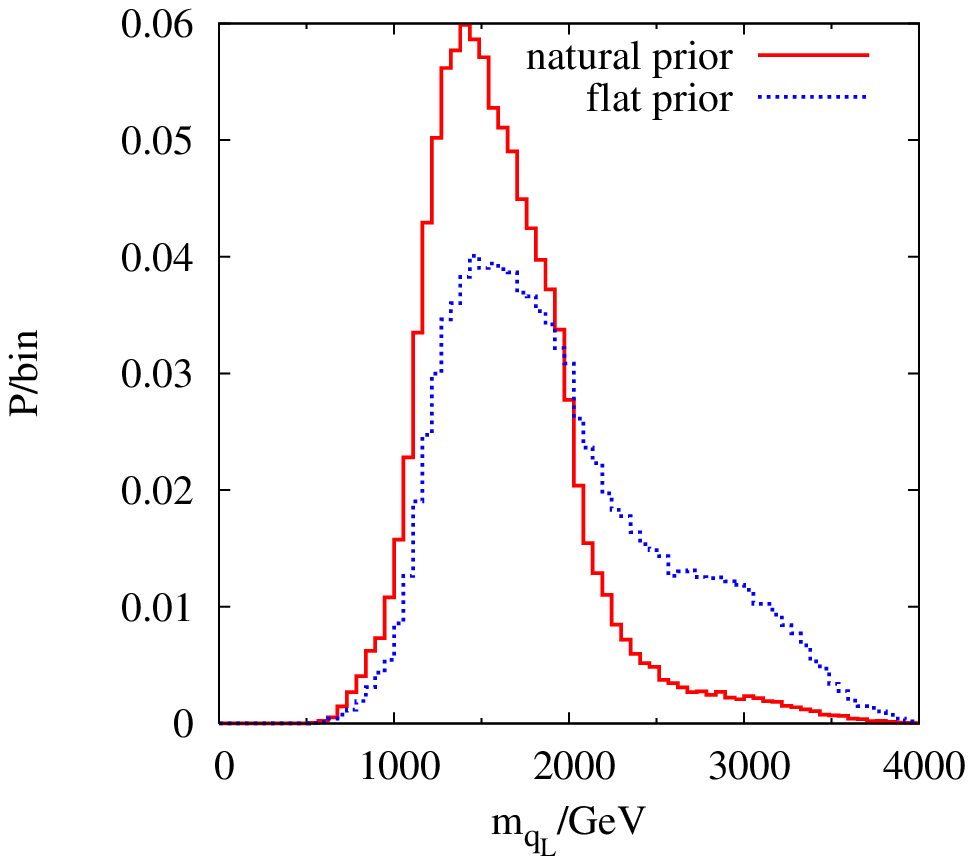}{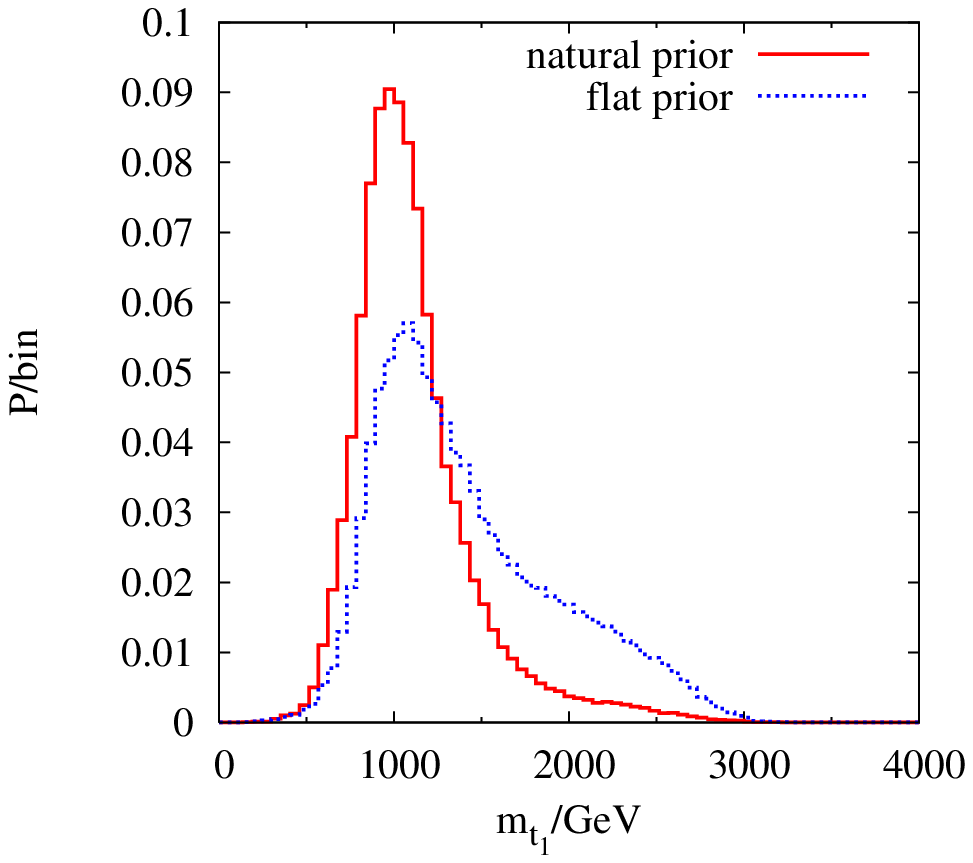}
\caption{Comparison of mass probability distributions for various sparticles with (``natural prior'') and without (``flat prior'') the naturalness prior.
Each histogram has been normalised to an integrated probability of 1. 
The plots show distributions for, respectively, (a) the lightest neutralino mass, (b) the lightest chargino mass, (c) the gluino mass, (d) the right-handed selectron mass, (e) the left-handed squark mass and (f) the lightest stop mass.} 
 \label{fig:spMasses}
 \end{center}
 \end{figure}
We now turn to the probability distributions for some of the sparticle masses, which are shown in Fig.~\ref{fig:spMasses}. The naturalness prior skews every distribution toward somewhat lighter masses except for the right-handed selectron mass, shown in Fig.~\ref{fig:spMasses}d. Along with the other sleptons, this particle becomes heavier due to the enhanced probability for the $h^0-$pole region, at large values of $m_0$. The squarks in Fig.~\ref{fig:spMasses}e do not become heavier because their mass is dominated by renormalisation group contributions from the gluino mass, which decreases with the naturalness prior, as can be seen in Fig.~\ref{fig:spMasses}c. Enhanced probabilities for very low values of $m_{\chi_1^0}$ and $m_{\chi_1^\pm}$ are particularly welcome because they enable a future International Linear Collider (ILC) with only 500 GeV centre of mass energy to produce them. For a 500 GeV machine, $\chi_1^0 \chi_1^0 (\chi_1^\pm \chi_1^\pm)$ thresholds can be reached with probabilities of 0.70,0.33 respectively, assuming the naturalness prior.
For an 800 GeV machine, the probabilities become 0.93,0.58 respectively. Of course, we should have some idea of the masses of these particles from initial data at the Large Hadron Collider (LHC). 
In order to measure sleptons by direct Drell-Yan pair production at the LHC, the sleptons must be fairly light: the cross-section $\sigma \sim 1$ fb$^{-1}$ for slepton masses of around 400 GeV would not provide many signal events and the $WW$ fusion rates are negligible because of cancellation with Bremsstrahlung processes~\cite{wwFusion}. The chance of directly observing sleptons: $m_{e_R}<400$ GeV, is 9$\%$ (8$\%$) for the flat and naturalness priors respectively.
From Fig.~\ref{fig:spMasses}, 95$\%$ C.L. upper bounds on the relevant sparticle masses are derived as shown in Table~\ref{tab:conf}. Caution must be exercised when interpreting upper confidence levels for the scalars: it is possible that the maximum value taken for $m_0=2000$ GeV is responsible for the upper bound. Presumably, a tail would extend to much higher scalar masses if the range of $m_0$ were increased. However, the fit itself prefers lighter gaugino masses and so they do not have this caveat. From the table, we see that the skew towards lighter masses is somewhat moderate. In either case, the bounds on the strongly interacting particles imply a good chance of SUSY discovery at the LHC~\cite{ATLASTDR}.
\begin{figure}
\begin{center}
\unitlength=1in
\begin{picture}(4,3)
\put(0,0){\epsfig{file=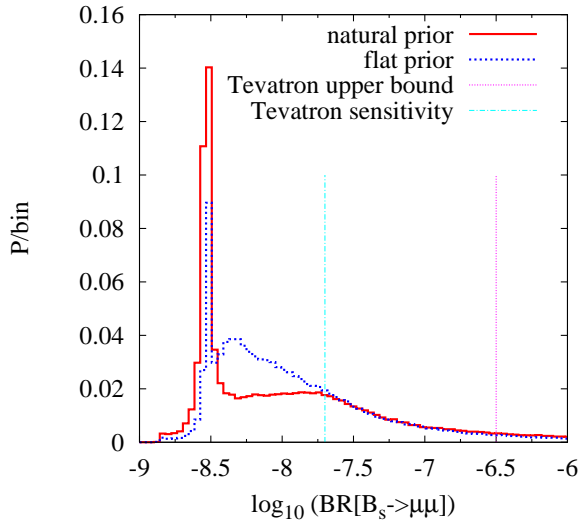, width=4in}}
\end{picture}
\caption{Comparison of probability distributions of the branching ratio of $B_s \rightarrow \mu^+ \mu^-$ with and without a naturalness prior. 
Each individual histogram has been normalised to an integrated probability of 
1.}  
 \label{fig:bsmumu}
 \end{center}
 \end{figure}

In Fig.~\ref{fig:bsmumu}, we compare the probability distribution of $BR(B_s \rightarrow \mu^+ \mu^-)$ as predicted by {\tt micrOMEGAs1.3.6} for the naturalness and flat priors.  
The range of predicted branching ratios is consistent with the range found by a random scan of unconstrained MSSM parameter space in Ref.~\cite{Dedes:2004yc}. The spike at low branching ratios is mostly due to the light higgs pole region's contribution and is enhanced for the natural prior. The current combined CDF/D0 95$\%$ C.L. limit~\cite{Acosta:2004xj}\footnote{There
  are newer preliminary CDF/D0 bounds~\cite{Dagu}, for example CDF(D0) have non-combined 
  95$\%$ C.L. limits of 2.0(3.0)$\times 10^{-7}$ respectively.}  $BR(B_s \rightarrow \mu^+ \mu^-)>3.4\times 10^{-7}$ is displayed by the right-hand vertical line, only ruling out a very small amount of parameter space. On the other hand, the expected future sensitivity of the Tevatron with 8 fb$^{-1}$ of integrated luminosity is $2 \times 10^{-8}$~\cite{future}, shown by the left hand vertical line. This would cover 32$\%$ of the integrated probability for either the flat or the natural priors. 

\section{Summary}
By employing a simple MCMC, we have investigated the effect of a naturalness
prior on CMSSM combined fits. The prior has a significant effect upon the
fits, indicating that current data are either not sufficiently precise or
not numerous enough to 
render posterior probabilities insensitive to the prior.
Our results bear some similarity to those of ref.~\cite{Giusti:1998gz}, which
also effectively used a naturalness prior, but did not use a likelihood, only
LEP 1,2 exclusion limits. However, features due to the dominant dark matter
annihilation mechanism are prominent in our results.
The naturalness prior enhances the posterior probability of the light CP-even higgs annihilation region of CMSSM parameter space, which comes back inside the 95$\%$ C.L. The relative probabilities of $A^0-$pole and stau co-annihilation regions decrease. Probability distributions for sparticle masses are skewed somewhat towards lower values, except for the sleptons, which show a slight skew towards heavier masses. The naturalness prior implies an enhanced probability for the lightest neutralinos and charginos to be accessible at a future linear collider. The prospects for evidence of $B_s \rightarrow \mu^+ \mu^-$ at the Tevatron remain good at 32$\%$ whichever prior is used for 8 fb$^{-1}$ of integrated luminosity. 
Rather light CP-even Higgs masses are preferred, the 95$\%$ upper C.L. limit
being only 120 GeV. This is unfortunately a somewhat difficult regime for
$h^0$ discovery by the LHC, and implies that several years of data-taking will
be required~\cite{ATLASTDR}. Ref.~\cite{gunion} found that fine-tunings of at
least 180 are required by the LEP2 bound $m_{h^0}>114$ GeV, $\tan
\beta=10$. Our results indicate fine-tunings in the range of 20-30 are still
feasible. 
The results in the present paper include wider variations of more
parameters (particularly in $A_0$), but the dominant difference to
Ref.~\cite{gunion} is that here we allow a 3 GeV theoretical error upon the
$m_{h^0}$ prediction, which extends the fine-tuning range of the fitted points
downwards. The random scan of Ref.~\cite{gunion} was also very sparse in the
parameters and so there is a distinct possibility of just not finding the less
fine-tuned points.

The amount of skew of the probability distributions will be modified by changing the prior assumption. For example, choosing a prior that is proportional to $1/c^2$ instead of $1/c$ (thus effectively counting the naturalness twice rather than just once) would produce more skew in the distributions. 

There are many possible future directions for the use of MCMC algorithms in fits to supersymmetric models. One obvious direction is the possibility of considering different supersymmetry breaking models, for example relaxing some of the universality assumed in the CMSSM\@. Another possible direction would be to include electroweak observables in the fit. Yet another would be to explore the $\mu<0$ part of parameter space and work out its probability normalisation with respect to the $\mu>0$ part. The MCMC procedure carried out here was at the limit of CPU capabilities available to the author, but finding a more efficient algorithm could allow for an enlargement of the parameter space considered. In particular, going to high values of $m_0$ would be useful in order to fully investigate the focus point regime~\cite{leszek}. 

 \section*{Acknowledgements}
 We would like to thank G G Ross for providing the idea and motivation for this  
 project, to CERN for hosting the computer facilities used and to M Drees for 
 an argument about the priors. 
 This work has been partially supported by PPARC.

\appendix
\section{Interpretation of the prior}
We are aware that readers that are unfamiliar with Bayesian statistics might feel uncomfortable at the appearance of the admittedly subjective priors (see for example, a footnote on page 6 of Ref.~\cite{Drees:2005bx}). A Bayesian viewpoint is that we are allowing the data to modify our uncertainty in some parameter and that probability is a measurement of our uncertainty. For example, one might estimate the chances of throwing a 6 on a straight cubic die to be 1 in 6, but knowledge of the position, the speed and the initial configuration of the die etc.\ would allow us to calculate the spin and trajectory of the die and so to modify the probability of throwing a 6.

Solace may be taken from the fact that the more precise data are, the more insensitive the posteriors will be to the priors. For instance, the data on the electron mass $m_e$ is so precise that different priors proportional to, say, $\ln(m_e)$, $e^{-m_e^2}$, $1/m_e^2 \ldots$ will result in almost identical posterior distributions for $m_e$. 
Choosing an extreme prior such as a delta-function for $m_e$ at 1 TeV, will result in a posterior delta function also at 1 TeV and it appears that many feel that this renders the Bayesian approach followed in this paper suspect.
Usually, the sub-text to this argument is the assumption that the prior is some probability distribution set by nature. However, if we interpret the prior as a measure of {\em our} uncertainty on $m_e$, setting it to be a delta function is equivalent to stating that we are absolutely sure of its value and we are not going to let data give us any information about it.

\end{document}